\documentclass[8.5pt,twoside,twocolumn]{article}
\usepackage[super,sort&compress,comma]{natbib} 
\usepackage[version=3]{mhchem}
\usepackage{balance}
\usepackage{times,mathptm}
\usepackage{secsty}
\usepackage{graphicx} 
\usepackage{lastpage}
\usepackage[format=plain,justification=raggedright,singlelinecheck=false,font=small,labelfont=bf,labelsep=space]{caption} 
\usepackage{fancyhdr}
\usepackage{amssymb}
\usepackage{pgfplots}
\usepackage[caption=false]{subfig}

\usetikzlibrary{arrows, decorations.markings}

\tikzstyle{vecArrow} = [very thick, decoration={markings,mark=at position
   1 with {\arrow[thick]{open triangle 60}}},
   double distance=1.4pt, shorten >= 5.5pt,
   preaction = {decorate},
   postaction = {draw,line width=1.4pt, white,shorten >= 4.5pt}]
\tikzstyle{innerWhite} = [semithick, white,line width=1.4pt, shorten >= 4.5pt]

\begin{document}

\thispagestyle{plain}
\fancypagestyle{plain}{
\fancyhead[L]{\includegraphics[height=8pt]{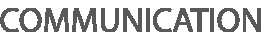}}
\fancyhead[C]{\hspace{-1cm}\includegraphics[height=20pt]{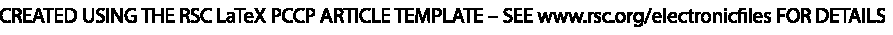}}
\fancyhead[R]{\hspace{10cm}\vspace{-0.25cm}\includegraphics[height=10pt]{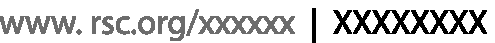}}
\renewcommand{\headrulewidth}{1pt}}
\renewcommand{\thefootnote}{\fnsymbol{footnote}}
\renewcommand\footnoterule{\vspace*{1pt}%
\hrule width 3.4in height 0.4pt \vspace*{5pt}} 
\setcounter{secnumdepth}{5}

\makeatletter 
\renewcommand\@biblabel[1]{#1}            
\renewcommand\@makefntext[1]%
{\noindent\makebox[0pt][r]{\@thefnmark\,}#1}
\makeatother 
\renewcommand{\figurename}{\small{Fig.}~}
\sectionfont{\large}
\subsectionfont{\normalsize} 

\fancyfoot{}
\fancyfoot[LO,RE]{\vspace{-7pt}\includegraphics[height=9pt]{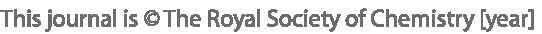}}
\fancyfoot[CO]{\vspace{-7.2pt}\hspace{12.2cm}\includegraphics{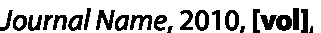}}
\fancyfoot[CE]{\vspace{-7.5pt}\hspace{-13.5cm}\includegraphics{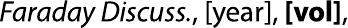}}
\fancyfoot[RO]{\footnotesize{\sffamily{1--\pageref{LastPage} ~\textbar  \hspace{2pt}\thepage}}}
\fancyfoot[LE]{\footnotesize{\sffamily{\thepage~\textbar\hspace{3.45cm} 1--\pageref{LastPage}}}}
\fancyhead{}
\renewcommand{\headrulewidth}{1pt} 
\renewcommand{\footrulewidth}{1pt}
\setlength{\arrayrulewidth}{1pt}
\setlength{\columnsep}{6.5mm}
\setlength\bibsep{1pt}

\twocolumn[
  \begin{@twocolumnfalse}
\noindent\LARGE{\textbf{Ultrafast Charge Separation and Nongeminate Electron-Hole Recombination in Organic Photovoltaics}}
\vspace{0.6cm}

\noindent\large{\textbf{Samuel L Smith$^{\ast}$\textit{$^{a}$} and Alex W Chin\textit{$^{a}$}}}\vspace{0.5cm}

\noindent\textit{\small{\textbf{Received Xth XXXXXXXXXX 20XX, Accepted Xth XXXXXXXXX 20XX\newline
First published on the web Xth XXXXXXXXXX 200X}}}

\noindent \textbf{\small{DOI: 10.1039/b000000x}}
 \end{@twocolumnfalse} \vspace{0.6cm}]

\noindent\textbf{The mechanism of electron-hole separation in organic solar cells is currently hotly debated. Recent experimental work suggests that these charges can separate on extremely short timescales ($<$100 fs). This can be understood in terms of delocalised transport within fullerene aggregates, which is thought to emerge on short timescales before vibronic relaxation induces polaron formation. However, in the optimal heterojunction morphology, electrons and holes will often re-encounter each other before reaching the electrodes. If such charges trap and cannot separate, then device efficiency will suffer. Here we extend the theory of ultrafast charge separation to incorporate polaron formation, and find that the same delocalised transport used to explain ultrafast charge separation can account for the suppression of nongeminate recombination in the best devices.}
\section*{}
\vspace{-1cm}
\footnotetext{\textit{$^{a}$Theory of Condensed Matter, Cavendish Laboratory, University of Cambridge, UK. E-mail: sls56@cam.ac.uk}}

The best solution-processed organic photovoltaic cells (OPVs) now exhibit efficiencies exceeding $9\%$ \cite{He2012a}. Devices consist of a nanostructured "heterojunction" morphology of intermixed electron donor and acceptor semiconductors \cite{Park2009}. Usually fullerene-derivatives are used as the electron acceptor. Photons are absorbed within the device generating tightly bound excitons. These excitons diffuse to interfaces between donor and acceptor semiconductor, where electron and hole can separate into free charges. However in order to separate, charges must overcome their mutual Coulomb attraction, which is an order of magnitude greater than thermal energies at room temperature \cite{Gelinas2011}. Experimentally, charge separation has been observed on ultrafast timescales ($<$100 fs) \cite{Bakulin2012, Grancini2012, Gelinas2014, Jailaubekov2012}. This observation is incompatible with conventional theories of charge transport in organic media, and new proposals have emerged \cite{Troisi2013,Caruso2012,Kaake2013,Bittner2014}. It has been proposed that ultrafast charge transfer is sustained by spatially coherent delocalised states, which arise on short timescales before molecular vibrations can respond to the presence of charges \cite{Gelinas2014,Bakulin2012, Tamura2013}. It is assumed that more localised polarons will form on longer timescales, although \textit{Bakulin et al.} found that delocalised states could be repopulated at late times by an infrared pulse \cite{Bakulin2012}. Electrons and holes remaining in close proximity at long times are thought to trap into bound charge transfer (CT) states at the interface, while separated charges are free to generate a photocurrent.

Electron hole pairs generate a dipolar electric field as they separate, this field can be observed as a Stark shift in the optical spectra of neighboring molecules. Using this signature, \textit{G$\acute{e}$linas et al.} directly observed the separation of charges on femtosecond timescales \cite{Gelinas2014}. They found that electron and hole separated by a few nanometres within just 40 fs, but that this was only observed in devices containing nanoscale aggregates of the fullerene-derivative electron acceptor PC$_{71}$BM. Alongside this experiment, we presented a simple phenomenological model of ultrafast charge separation through delocalised states of small acceptor crystals. This model has since been supported by more detailed modeling of PCBM crystallites \cite{Savoie2014}. Our central proposition states that in order for ultrafast charge separation to occur, the effective bandwidth of the crystallite LUMO should be similar in magnitude to the electrostatic binding energy of electron and hole across the interface.

The current theoretical model is not complete, since it only describes charge transport within a few hundred femtoseconds of exciton dissociation. Once charges are free, they will diffuse through the device. However, in the heterojunction morphology many electrons and holes will re-encounter each other before reaching the electrodes. This may lead to re-trapping and nongeminate exciton recombination, lowering the device efficiency \cite{Street2010, Credgington2011, Credgington2012}. In efficient devices, either re-trapping must be suppressed or the trapped CT states that form must themselves be able to separate into free charges. In a recent experiment, \textit{Rao et al.} noted the absence of nongeminate triplet excitons at open circuit in an efficient PIDT-PhanQ:PC$_{60}$BM device \cite{Rao2013}. Since three quarters of nongeminate CT states should have triplet character, they concluded that such CT states were able to separate long after exciton dissociation first occurs, thus avoiding the formation of triplet excitons. To explain these observations and build a complete description of charge separation on both femtosecond and nanosecond timescales, we extend our description of ultrafast charge separation to take account of vibronic relaxation. 

First we briefly reiterate our model of ultrafast charge separation \cite{Gelinas2014}. This model is illustrated in figure 1a, in figure 1b we illustrate our parallel theory of trapped pair separation at late times (developed below). We assume that ultrafast charge separation occurs when an exciton reaches an interface between donor molecules and a small acceptor crystallite. This crystallite is modeled by an FCC lattice of localised single electron energy levels, which are coherently coupled to their nearest neighbours. We take a Gaussian distribution of site energies to introduce disorder. The electronic eigenstates of this crystallite are delocalised standing waves, with bandwidth B. Then we introduce a Coulomb well surrounding a donor site adjacent to one face of the crystal, this well models the hole left behind after an electron is injected into the crystallite. The Hamiltonian,
\begin{equation}
H_S  =  \sum_i E_i |i\rangle \langle i| - \sum_{n.n.} J |i\rangle \langle j|. \label{eq:3} \\
\end{equation}
$E_i = \sigma_i - \frac{e^2}{4 \pi \epsilon_0 \epsilon_r r_i}.$ $\sigma_i$ represents the Gaussian disorder on each site, $\epsilon_r$ labels the dielectric constant, and $r_i$ labels the distance between the i$^{th}$ acceptor site and the hole. $J$ labels the transfer integral between nearest neighbours, and the bandwidth B $= 16J$. The maximum depth of the Coulomb well within the acceptor lattice, W $= e^2/4\pi \epsilon_0 \epsilon_r r_1$. If the bandwidth B is much smaller than W, then the electronic eigenstates at the interface will localise and ultrafast charge transport will not occur. However if $B \gtrsim W$, then a set of band-states will survive, delocalised across the entire crystallite. Resonant coupling between the incoming exciton and these delocalised states can drive ultrafast charge separation.

A detailed DFT study of the popular electron acceptor PC$_{71}$BM was recently published, which supports this picture \cite{Savoie2014}. Despite the Coulomb well induced by the hole, nanoscale crystallites were found to exhibit delocalised states near the interface, able to support resonant coupling with incoming excitons. The naive LUMO bandwidth is $\sim$0.15 eV, rather smaller than the likely Coulomb well depth of $\sim$0.3 eV. However, PC$_{71}$BM exhibits three closely spaced low lying molecular orbitals; incorporating all three low lying bands enhances the effective bandwidth to $\sim$0.4 eV. This ensures the existence of electronic states near the interface resonant with states in the bulk.

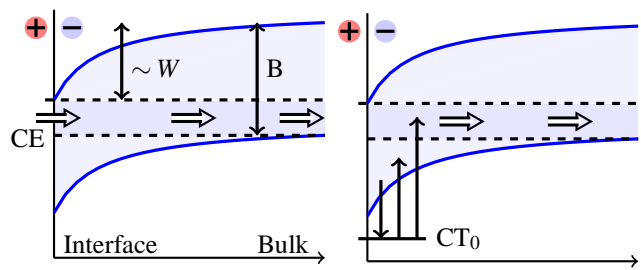
\begin{figure}[t]
\subfloat{\begin{tikzpicture}[scale=0.6]
	\fill[blue!10!white] (0,3.714) rectangle (6,4.48);
	\fill[blue!5!white, domain = 0:6] plot (\x, {6.5-2/(\x+1)});
	\fill[blue!5!white] (0,4.5) -- (6,4.5) -- (6,6.214);	
	\fill[blue!5!white, domain = 6:0] (0,3.714) -- plot (\x, {4-2/(\x+1)});

	\fill[red!50!white] (-0.4,6.1) circle (0.25);	
	\fill[blue!20!white] (0.4,6.1) circle (0.25);
	\draw[very thick] (-0.6,6.1) -- (-0.2,6.1);
	\draw[very thick] (-0.4,5.9) -- (-0.4,6.3);
	\draw[very thick] (0.2,6.1) -- (0.6,6.1);

	\draw[thick,->] (0,1) -- (6,1);
	\draw[thick] (0,1) -- (0,6.5);

	\draw[color=blue, domain = 0:6, very thick] plot (\x, {6.5 - (2/(\x+1)^1)});
	\draw[color=blue, domain = 0:6, very thick] plot (\x,{4 - (2/(\x+1)^1)});

	\draw[<->, very thick] (4.5,3.714) -- (4.5,6.214);
	\node at (4.5, 5.2) [right] {B};

	\draw[<->, very thick] (1.5,4.5) -- (1.5,6.214);
	\node at (1.5, 5.2) [right] {$\sim W$};

	\draw[dashed, very thick] (0,3.714) -- (6,3.714);
	\draw[dashed, very thick] (-0.2,4.5) -- (6,4.5);
	
	\node at (0,1.3) [right] {Interface};
	\node at (4.3,1.3) [right] {Bulk};
	\node at (0,3.654) [left] {CE};

	\draw[vecArrow] (-0.4,4.1) to (0.6, 4.1);
	\draw[vecArrow] (2.6,4.1) to (3.6,4.1);
	\draw[vecArrow] (5,4.1) to (6, 4.1);
\end{tikzpicture}}\hspace{0.2mm}
\subfloat{\begin{tikzpicture}[scale=0.6]
\fill[blue!10!white] (0,3.714) rectangle (6,4.48);
	\fill[blue!5!white, domain = 0:6] plot (\x, {6.5-2/(\x+1)});
	\fill[blue!5!white] (0,4.5) -- (6,4.5) -- (6,6.214);	
	\fill[blue!5!white, domain = 6:0] (0,3.714) -- plot (\x, {4-2/(\x+1)});

	\fill[red!50!white] (-0.4,6.1) circle (0.25);	
	\fill[blue!20!white] (0.4,6.1) circle (0.25);
	\draw[very thick] (-0.6,6.1) -- (-0.2,6.1);
	\draw[very thick] (-0.4,5.9) -- (-0.4,6.3);
	\draw[very thick] (0.2,6.1) -- (0.6,6.1);

	\draw[thick,->] (0,1) -- (6,1);
	\draw[thick] (0,1) -- (0,6.5);

	\draw[color=blue, domain = 0:6, very thick] plot (\x, {6.5 - (2/(\x+1)^1)});
	\draw[color=blue, domain = 0:6, very thick] plot (\x,{4 - (2/(\x+1)^1)});

	\draw[dashed, very thick] (0,3.714) -- (6,3.714);
	\draw[dashed, very thick] (-0.2,4.5) -- (6,4.5);

	\draw[very thick] (-0.2,1.5) -- (1.3,1.5) node[right] {CT$_0$};
	\draw[very thick, ->] (0.3,2.8) -- (0.3,1.5);
	\draw[very thick, <-] (0.7,3.3) -- (0.7,1.5);
	\draw[very thick, ->] (1.1,1.5) -- (1.1,4.2);

	\draw[vecArrow] (1.6,4.1) to (2.6,4.1);
	\draw[vecArrow] (4,4.1) to (5, 4.1);
\end{tikzpicture}}
\caption{Left (a): Electronic eigenstates on ultrafast timescales. If W $\lesssim$ B, then a set of delocalised states will survive with signiﬁcant weight near the interface. These states enable ultrafast charge separation. Right (b): On longer timescales vibronic relaxation lowers the energy of the CT$_0$ state, while leaving higher lying states unchanged. Thermal fluctuations promote the trapped electron into higher lying states, excitation into the delocalised states above CE can enable charge separation.}
\end{figure}

The account above is only intended to describe particle dynamics within the first hundred femtoseconds after exciton dissociation. To consider the electronic eigenstates of nongeminate trapped CT states, formed on long timescales, we will have to incorporate vibronic relaxation. We couple each lattice site to an effective molecular vibration \cite{Holstein1959, Troisi2011},
\begin{eqnarray}
H & = & H_S + H_B + H_I,\label{eq:1} \\
H_B & = & \omega \sum_i a_i^{\dagger}a_i, \\
H_I & = & g \sum_i (a_i + a_i^{\dagger})|i\rangle \langle i|. 
\end{eqnarray}
We express the electronic subsystem in terms of the general wavefunction $ |\psi\rangle = \sum_i C_i |i\rangle $, and the vibrations in terms of position and momentum operators $X_i(t) = \langle a_i + a_i^{\dagger} \rangle$, $P_i(t) = \langle a_i - a_i^{\dagger} \rangle$. These obey the Heisenberg equation of motion $\dot{O} = i[H,O]$. We assume that vibrations oscillate slowly compared to electronic modes. This allows us to treat the vibrations semi-classically, making the approximation $H_I \to g \sum_i X_i|i\rangle \langle i|.$ 

Thus far we have neglected damping in the vibrational mode, which is necessary to reach the relaxed lowest energy CT$_0$ state. This can easily be included phenomenological into the equations of motion,
\begin{eqnarray}
\dot{X_i} & = & -i \omega P_i, \\
\dot{P_i} & = & -i(\omega X_i + 2g |C_i|^2) - \gamma \dot{P_i}.
\end{eqnarray}
Damping will drive both $P_i$ and $\dot{P_i}$ to zero. Thus in the relaxed CT$_0$ state, $X_i = -2g|C_i^0|^2/\omega$. Inserting this result into equation \ref{eq:1}, we obtain an effective non-linear Hamiltonian for the electronic eigenstates of the relaxed, maximally trapped CT states at long times \cite{Emin},
\begin{equation}
H_{\text{eff}} = \sum_i (E_i - \Delta |C_i^0|^2)  |i\rangle \langle i| + \sum_{n.n.} J|i\rangle \langle j| .
\label{eq:2}
\end{equation}
The reorganization energy $\Delta = g^2/\omega$ \footnote{In this simple model, the energy penalty arising from $H_B$ is precisely half the polaronic stabalisation arising from $H_I$. The relevant reorganization energy here corresponds to the upper potential energy surface of Marcus theory; if $H_B$ is identical for occupied and unoccupied electronic states then $\Delta = \Delta_{Marcus}/2$.}. This equation can be simply understood. At long times the hole is assumed to lie next to an acceptor crystallite, while the electron lies in the lowest available eigenstate of the acceptor lattice. The vibrational reorganization on each electronic site is proportional to the charge density on that site. This lowest energy eigenstate $|\psi_0\rangle = \sum_i C_i^0 |i\rangle$ can be found by a simple iterative procedure. Since the vibrational modes are assumed slow compared to electronic motion, the instantaneously accessible higher energy eigenstates can be found by holding the vibration reorganization on each site fixed and solving $H_{\text{eff}}$, which is now linear, for the higher lying states.

Vibronic relaxation will tend to localise the CT$_0$ eigenstate near the interface. If the electron is localised on a single site then the corresponding site energy will be reduced by $\Delta$. This will provide a significant barrier to charge separation. However if the couplings between neighboring lattice sites are sufficiently strong to preserve a delocalised CT$_0$, then the corresponding reorganization energies of the relevant lattice sites will be reduced; lowering the barrier to charge separation. Additionally, delocalisation will have decreased the exciton recombination rate.

We now investigate the properties of the relaxed CT$_0$ state numerically. We take a dielectric constant of 3, a donor-acceptor nearest neighbour separation $r_1 =$ 1.5 nm (implying W $=$ 0.32 eV), and a static disorder standard deviation of 50 meV. The FCC lattice constant is 1.5 nm, and the acceptor crystallite comprises 4$^3$ unit cells. Finally we fix the nearest neighbour coupling $J = 25$ meV (implying a bandwidth B $=$ 0.4eV). 
\begin{figure}[t]
\subfloat{\includegraphics[scale = 0.5]{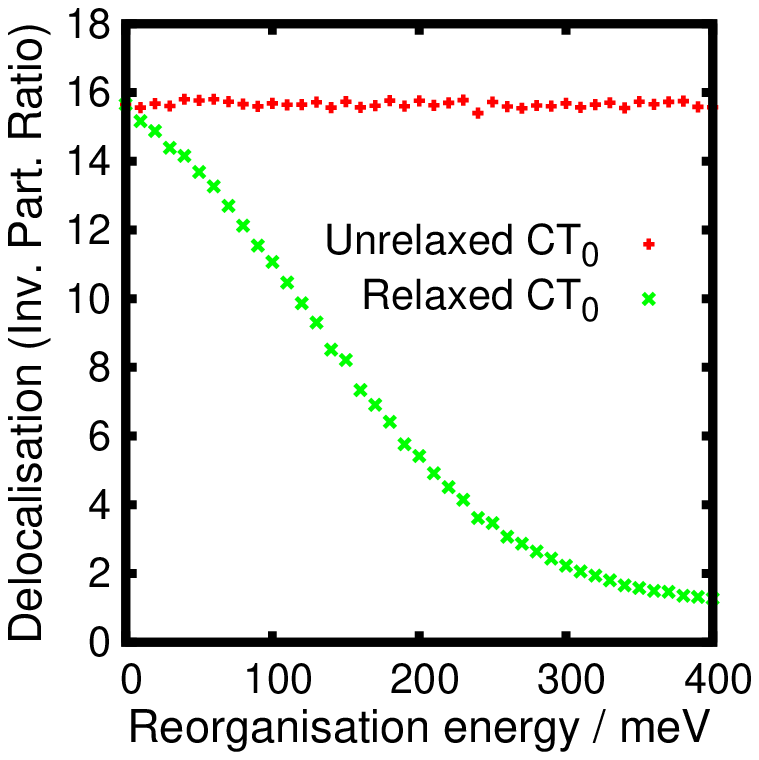}}\hspace{0.2mm}
\subfloat{\includegraphics[scale = 0.5]{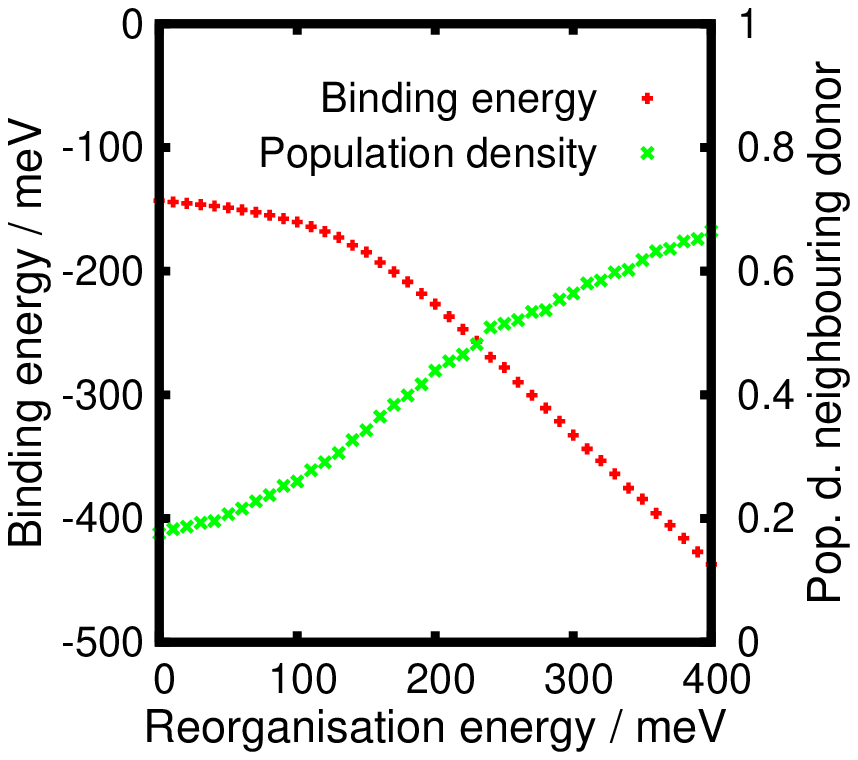}} \\
\subfloat{\includegraphics[scale = 0.5]{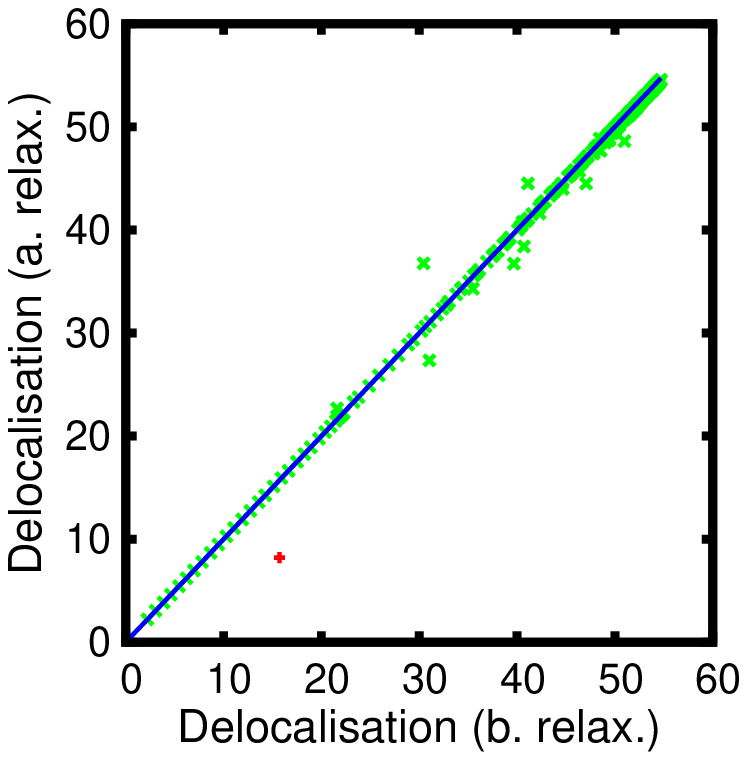}}\hspace{0.2mm}
\subfloat{\includegraphics[scale = 0.5]{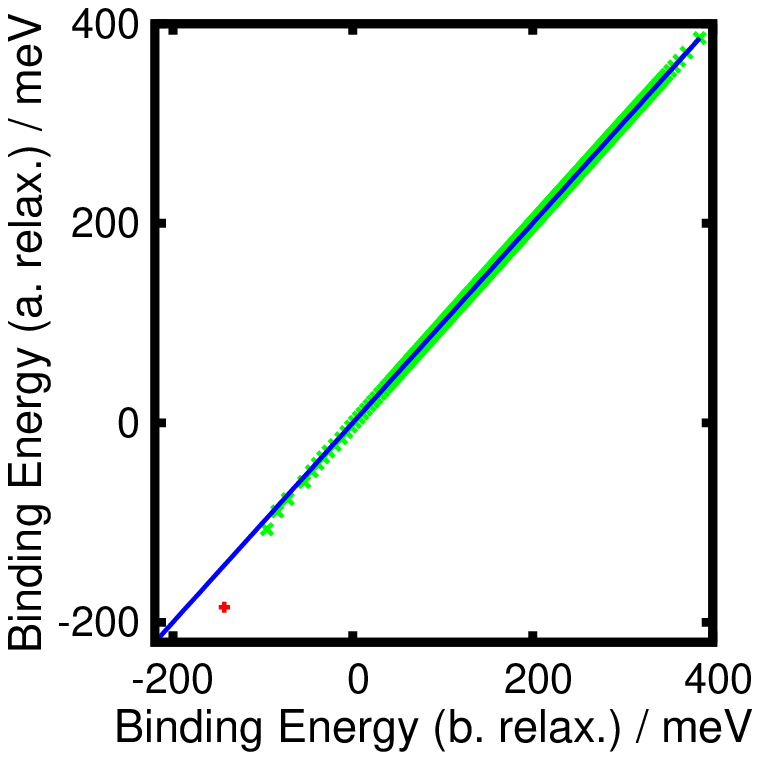}}
\caption{Top left (a): Delocalisation of the lowest electronic eigenstate with and without vibronic relaxation. At early times, before relaxation occurs, the state is delocalised. At long times the localisation depends on the reorganization energy. Top right (b): Binding energy of the charge pair, and population density neighboring the donor. Bottom panel: Delocalisation and binding energy of all electronic eigenstates before and after relaxation. States are labelled by their energy, and then plotted against each other. The CT$_0$ state (red) is more localised and more strongly bound after relaxation occurs. Higher lying states (green), when averaged over disorder runs, are not significantly affected by the relaxation process. Consequently these states lie on the line $y=x$ (blue).}
\end{figure}

In figure 2a we exhibit the mean delocalisation of the CT$_0$ eigenstate as $\Delta$ is varied between 0 and 0.4 eV. For reference, and to illustrate the simulation error, we also include the delocalisation of the early time unrelaxed lowest eigenstate of equation \ref{eq:3}. Each data point is averaged over 10000 runs, and the delocalisation is measured via the inverse participation ratio\footnote[2]{The inverse participation ratio, IPR $=  1/(\sum_i|C_i|^4)$, where the sum runs over all lattice sites.}. As discussed in our earlier paper, at early times before vibrational relaxation can occur even the lowest eigenstate of the acceptor lattice is delocalised over many sites. This occurs despite the presence of a deep Coulomb well near the interface. By contrast, when $\Delta$ is large, the relaxed CT$_0$ eigenstate which forms on long timescales is fully localised. In this limit nongeminate trapped electron hole pairs would be unlikely to separate. As $\Delta$ is reduced below 0.3 eV however, the relaxed CT$_0$ state begins to delocalise.

To further explore this transition, in figure 2b we present the binding energy of the electron in the relaxed CT$_0$ state as well as the population fraction of this state which lies on the acceptor site neighboring the donor. The binding energy is measured from the bottom of the LUMO band of bulk acceptor crystal, $E_B = E+12J$\footnote[3]{Note that FCC lattices show an unusual band structure. In the infinite bulk crystal without disorder, there are equal numbers of eigenstates above and below the isolated site energy $E=0$. However the bandwidth, which extends from $-12J$ to $+4J$, is not symmetric.}. The binding energy of the CT$_0$ is little affected until $\Delta >$ 100 meV, after which it rises steeply. Meanwhile as $\Delta$ is reduced the population density neighboring the donor site falls, enhancing the CT$_0$ state lifetime and giving the charge pair more opportunity to separate. Typical values for the reorganization energy of $\pi$-conjugated organic molecules lie in the range 0.1-0.3 eV \cite{Coropceanu2007} \footnote[4]{There is some uncertainty over the intramolecular reorganization energy of fullerene derivatives. \textit{Kwiatkowski et al.} computed a theoretical Marcus value of 0.13 eV for C$_{60}$ \cite{Kwiatkowski2009}, and \textit{Cheung et al.} computed a similar value of 0.14 eV for PC$_{60}$BM \cite{Cheung2010}. These results include both upper and lower potential surfaces, suggesting a reorganisation energy here of $\sim$70 meV. However \textit{Savoie et al.} directly compute the reorganisation energy of the upper potential surface (the anion) of PC$_{60}$BM and obtain a much smaller value of just 15 meV \cite{Savoie2014}.}; which suggests that even the fully relaxed CT$_0$ state may remain delocalised over several sites near the interface, if the bandwidth is sufficiently large.

Thus far we have compared the lowest eigenstate of the acceptor crystallite at early and late times. However, so long as the molecular modes can be treated as slow, we may also compare the relaxed and unrelaxed higher lying electronic eigenstates. In the bottom panel of figure 2, we take $\Delta =$ 0.15 eV, and directly plot the binding energy and delocalisation of the relaxed eigenstates against their early time counterparts. We observe that, while relaxation lowers the energy and delocalisation of the lowest eigenstate (CT$_0$), it has much less effect on the higher lying states. This is easily understood, since the vibrational reorganization is determined by the CT$_0$ state density. It explains why \textit{Bakulin et al.} were able to optically re-excite the CT$_0$ state to the delocalised band states typically observed immediately after exciton dissociation \cite{Bakulin2012}.

After vibrational relaxation has occurred, thermal fluctuations in the vibrational modes attached to each lattice site drive spontaneous transitions from the relaxed CT$_0$ state to higher lying states. Applying time dependent perturbation theory and assuming that vibrational fluctuations are in thermal equilibrium, the rate of such transitions is given by
\begin{equation}
R_{0\to a} = 2\pi  \frac{J(E_a)}{e^{(E_a-E_0)/k_BT} - 1} \sum_{i} |C_i^a C_i^0|^2. \label{eq:rate}
\end{equation}
The spectral density of a single, over-damped vibrational mode, $J(E) = \Delta E \gamma/(E^2 + \hbar^2 \gamma^2)$ \cite{Mukamel1995}. In figure 3a we take a damping timescale $2\pi/\gamma = 100$fs, and we plot the "escape timescale" as a function of the reorganization energy\footnote[5]{The probability distribution of rates in the presence of static disorder is highly skewed, due to the exponential factor in equation \ref{eq:rate}. Therefore we take the average over the binding energy and the overlap function, before calculating the rate using these mean values.}. This denotes the typical time required for an electron to be excited, from the maximally trapped CT$_0$ state, to one of the unbound eigenstates with binding energy $E_B > 0$. As demonstrated above, these are the same states which drive ultrafast charge separation; each such event presents an opportunity for electron and hole to separate.

We focus first on the green curve at 300K. Since transitions are driven by thermal fluctuations in the vibrational modes, the escape timescale diverges as $\Delta \to 0$. It exhibits a broad minimum at $\Delta \approx$ 70 meV and rises rapidly for $\Delta >$ 200meV. The electron-hole recombination timescale is thought to be a few nanoseconds \cite{Deibel2010}; therefore, so long as $\Delta <$ 200meV, the electron will experience several opportunities to escape the hole before recombination occurs. We predict that an electron acceptor with sufficiently broad bandwidth ($\approx 0.4 eV$) may not only support ultrafast charge separation, but can also assist the thermal separation of electrons and holes on timescales sufficiently fast to suppress nongeminate recombination. This unified description of ultrafast charge separation and the dissociation of trapped pairs was summarised in figure 1, both phenomena depend crucially on the formation of nanoscale acceptor crystallites ($\gtrsim$ 5$^3$ nm$^3$). Whereas ultrafast charge separation is temperature independent \cite{Gelinas2014}, the curves at 200/400K emphasize that the separation of trapped nongeminate pairs is highly temperature dependent. These predictions agree well with the observations of \textit{Rao et al.} \cite{Rao2013}.
\begin{figure}[t]
\subfloat{\includegraphics[scale = 0.5]{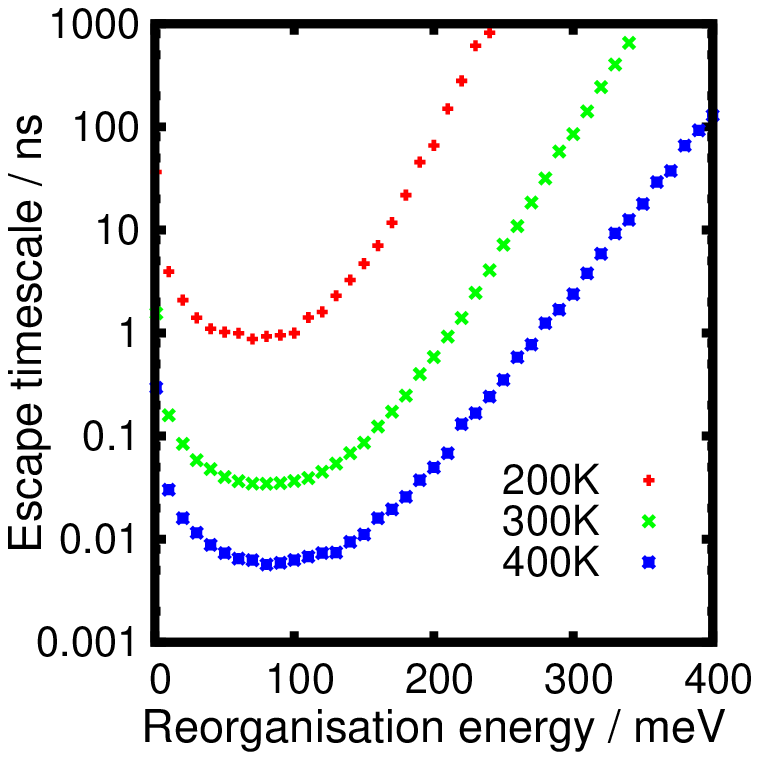}}\hspace{0.2mm}
\subfloat{\includegraphics[scale = 0.5]{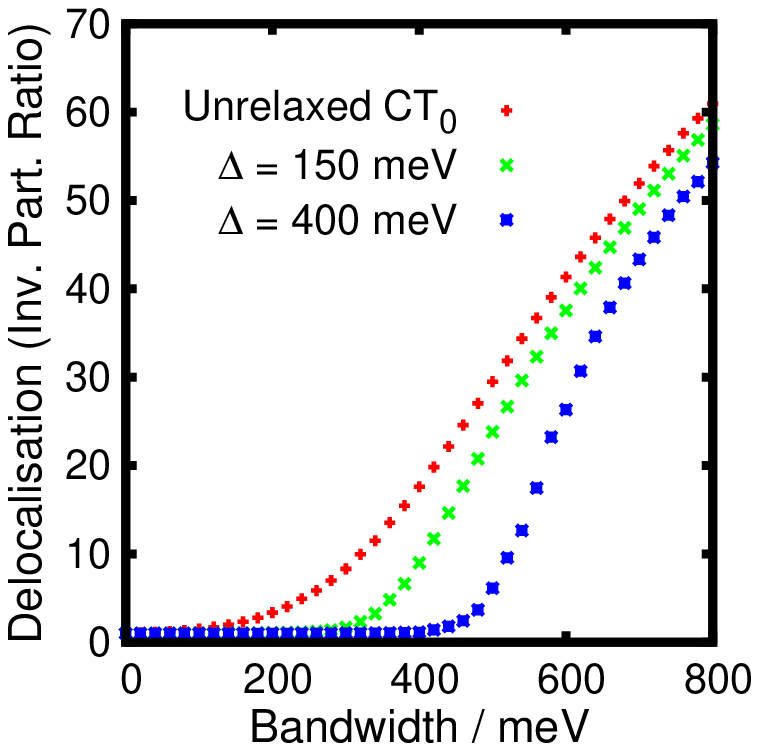}} \\
\subfloat{\includegraphics[scale = 0.5]{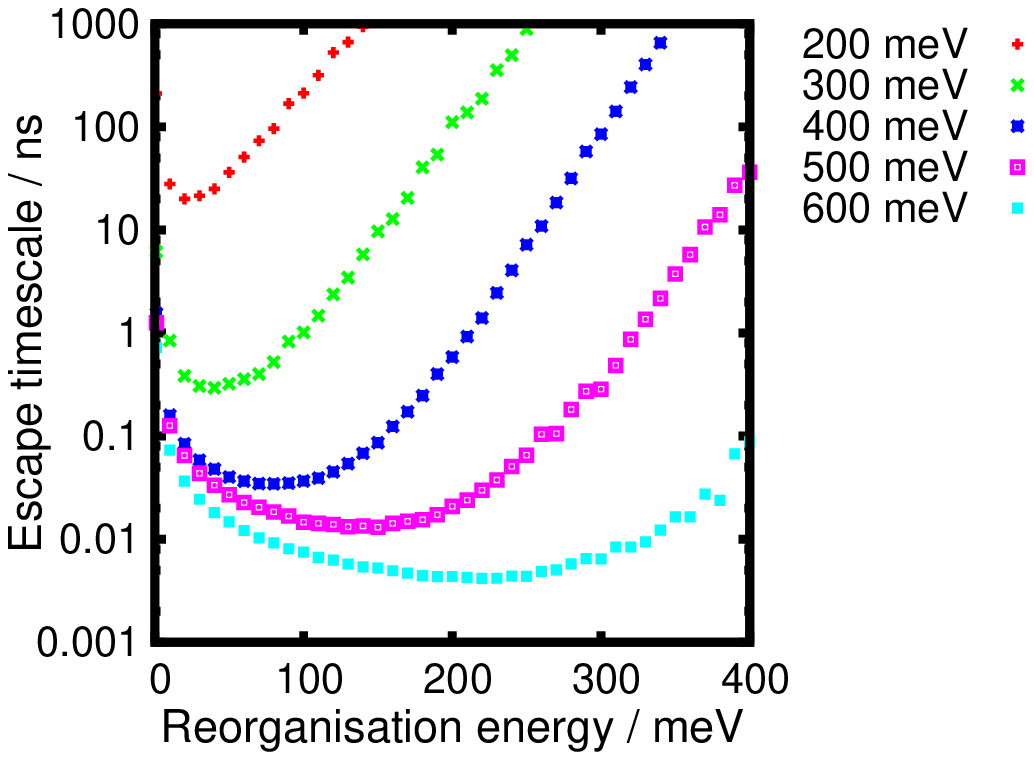}}
\caption{Top left (a): The transition timescale from the CT$_0$ state to higher lying, unbound states, at a range of temperatures. Top right (b): The delocalisation of the CT$_0$ state for $\Delta =$ 0, 150 and 400 meV, as a function of bandwidth. Bottom (c): The escape timescale at 300K for a range of bandwidths. All points (a-c) averaged over 1000 runs.}
\end{figure}

Thus far we have only considered an efficient device with relatively large bandwidth, $B =$ 0.4eV. In figure 3b we plot the delocalisation of the CT$_0$ state as the bandwidth is varied from 0 to 0.8 eV, for three different values of the reorganization energy. Delocalisation rises rapidly once the bandwidth is increased beyond a particular threshold; for moderate reorganization ($\Delta =$ 150meV) this threshold is determined by the Coulomb well depth W $=$ 320meV. Finally, in figure 3c we plot the escape timescale at 300K for a range of bandwidths. A bandwidth of at least 0.3eV appears necessary to suppress nongeminate recombination, and the resilience of a device to vibrational reorganization increases dramatically as the bandwidth rises.

As Troisi and co-workers have emphasized, organic crystals exhibit not only static energetic disorder, but also off diagonal disorder in the couplings between neighboring molecules \cite{Troisi2006,Troisi2011}. Also problematic is the assumption of a single broad LUMO band; as discussed earlier, the CT eigenstates of PCBM crystallites, the most popular electron acceptor, are formed from the mixing of three closely spaced narrow bands \cite{Savoie2014, Cheung2010}. In future work we hope to investigate these two effects in more detail, here we prefer to preserve a simple and intuitive picture of the acceptor crystallite.

We have not allowed any delocalisation of the hole. It is likely that in real systems the hole is partially delocalised along the donor polymer; this will lower the binding energy between electron and hole, reducing the acceptor bandwidth required to enable ultrafast charge separation and suppress exciton recombination \cite{Tamura2013}. Additionally, we have pessimistically assumed that the charge pair is able to relax into the maximally trapped CT$_0$ state described by equation \ref{eq:2}. In reality, the thermal fluctuations in the vibrational modes obstruct the relaxation process. Consequently, at sufficiently high temperatures a stable vibronically relaxed CT$_0$ state will never form. In this instance the eigenstates of the acceptor crystallite will resemble their early time counterparts at all times. To quantify this temperature, we note that the typical fluctuation scale of molecular vibrations on a lattice site is set by $k_BT \sim$ 25 meV at room temperature. Meanwhile the typical vibrational reorganization energy on a site is given by $\Delta/D$, where D is the delocalisation of the CT$_0$. Consequently, once D exceeds 5-10 sites, thermal fluctuations are similar in scale to the molecular reorganization, and the formation of a polaronic CT$_0$ state is likely to be inhibited. 

The discussion above has focused on the thermal seperation of nongeminate pairs, however the mechanism is equally applicable to the seperation of trapped geminate pairs. As such, these results may rationalise the apparent contradiction between ultrafast charge transport and the observations of \textit{Vandewal et al.} \cite{Vandewal2014a}; who found that sub-gap excitation of trapped CT states can efficiently generate free charges.  

In conclusion, there is significant experimental evidence that charge pairs are able to separate on ultrafast timescales at the interfaces between organic donor semiconductors and fullerene-derivative crystallites. This observation is best understood by treating the electronic eigenstates within these crystallites as delocalised. We have shown that simple models of this phenomenon can be extended to incorporate vibronic relaxation, which has previously been thought to localise these states. By contrast, here we show that the electronic eigenstates can remain delocalised and continue to assist the separation of trapped electron-hole pairs on long timescales. This work underlines the importance of crystallinity and nanoscale morphology for OPV performance.

We acknowledge funding from the Winton Programme for the Physics of Sustainability and we thank Richard H. Friend, Neil Greenham, Akshay Rao and Dan Credgington for helpful comments on the manuscript.



\footnotesize{
\bibliography{library2} 
\bibliographystyle{rsc} }

\end{document}